\documentclass{WileyMSP-template}

\usepackage{hyperref}
\usepackage[dvipsnames]{xcolor}
\usepackage{graphicx}% Include figure files
\usepackage{dcolumn}% Align table columns on decimal point
\usepackage{subfigure}
\usepackage{bm}% bold math
\usepackage[square, super]{natbib}
\usepackage{amsmath} 
\usepackage{soul}
\usepackage{ulem}

\begin{document}

\pagestyle{fancy}
\rhead{\includegraphics[width=2.5cm]{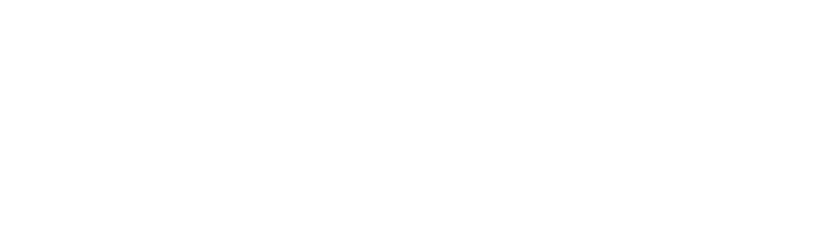}}

\title{Double helical plasmonic antennas}

\maketitle

\author{Aleksei Tsarapkin*}
\author{Luka Zurak}
\author{Krzysztof Maćkosz}
\author{Lorenz Löffler}
\author{Victor Deinhart}
\author{Ivo Utke}\\
\author{Thorsten Feichtner}
\author{Katja Höflich*}

\dedication{}

\begin{affiliations}
A. Tsarapkin, V. Deinhart, Dr. K. Höflich\\
Ferdinand-Braun-Institut (FBH), Gustav-Kirchhoff-Str. 4, D-12489 Berlin, Germany\\
Email Addresses: aleksei.tsarapkin@fbh-berlin.de, katja.hoeflich@fbh-berlin.de

L. Zurak, L. Löffler, Dr. T. Feichtner\\
Experimental Physics 5, University of Würzburg, Am Hubland, D-97074 Würzburg, Germany\\

V. Deinhart\\
Max Born Institute for Nonlinear Optics and Short Pulse Spectroscopy, Max-Born-Str. 2A, D-12489 Berlin, Germany\\

Dr. K. Maćkosz, Dr. I. Utke\\
Laboratory for Mechanics of Materials and Nanostructures, Empa -- Swiss Federal Laboratories for Materials Science and Technology, Feuerwerkerstrasse 39, CH-3602 Thun, Switzerland
\end{affiliations}


\begin{abstract}
Plasmonic double helical antennas are a means to funnel circularly polarized light down to the nanoscale. Here, an existing design tool for single helices is extended to the case of double helices and used to design antennas that combine large chiroptical interaction strength with highly directional light emission. Full-field numerical modeling underpins the design and provides additional insight into surface charge distributions and resonance widths. The helical antennas are fabricated by direct writing with a focused electron beam, a technique that is unrivaled in terms of spatial resolution and 3D shape fidelity. After the printing process, the structures are purified using ozone plasma at room temperature, resulting in the smallest continuous double helix antennas ever realized in gold. Fabricated antennas are studied regarding their polarization-dependent transmission behavior, which shows a large and broadband dissymmetry factor in the visible range. Since the polarization of light is an important tool for implementing logic functionality in photonic and quantum photonic devices, these helices are potential building blocks for future nanophotonic circuits, but also for chiral metamaterials or phase plates.
\end{abstract}

% Please make the first reference to a display item bold: \textbf{Figure 1}
% Do not abbreviate Figure, Equation, etc.; display items are always singular, i.e., Figure 1 and 2.
% Equations are always singular, i.e., Equation 1 and 2, and should be inserted using the {equation} environment, not as graphics
% Please do not use footnotes in the text, additional information can be added to the Reference list.

\section{\label{Introduction}Introduction}

The enhancement of light-matter interaction and the coherent manipulation of light at the nanoscale are key to photonic quantum technology, to enable the transition from bulky laboratory setups to miniaturized on-chip devices~\cite{Moody2022}.
This requires the integration of quantum functionalities onto photonic chips, i.e.~solid-state based qubits strongly coupled to a single optical mode populated with photons carrying the quantum information.
According to Purcell's formula~\cite{Purcell1946}, there are two fundamentally different approaches to enhance light-matter interaction, both of which have their own advantages and shortcomings.
Either the interaction time can be increased by minimizing losses in resonators, leading to extremely large quality factors $Q$~\cite{Vahala2003}, or the interaction probability can be increased by reducing the mode volume $V$ through resonant polariton-based near-field interactions in open cavities or antenna systems~\cite{Biagioni2012, Koenderink2017}.
The latter can be based on surface plasmon polaritons in metals arising from the collective motion of free electrons coupled to visible light~\cite{Maier2007} or other polaritonic material resonances such as exciton or phonon polaritons~\cite{Basov2016}.
Especially resonant plasmonic antennas allow to combine extremely high emitter decay rates due to a small mode volume with a relatively broadband operation due to their moderate Q factor~\cite{Koenderink2017}.
These antennas can be further designed to couple to specific radiating modes~\cite{Feichtner2017} to e.g.~achieve directional emission of quantum emitters~\cite{Taminiau2008}. 
In addition, chiral plasmonic components enable strongly enhanced chiroptical interactions~\cite{Hentschel2017, fernandez2016objects, Fernandez-Corbaton2023} to tailor the circular polarization state of the emitted light in a compact manner.
Since polarization encoding is one of the most popular methods used in photonic quantum protocols~\cite{Flamini2019}, the realization of such nanoscale components that improve quantum emission and additionally provide directional and polarization control is in high demand.

In RF-technology, direct access to (far-field) circular polarization is realized by helical antennas ~\cite{balanis2016antenna}. 
A plasmonic single turn helix -- substantially smaller than the wavelength of light -- can be regarded as a nearly perfect chiral dipole for its fundamental resonance~\cite{Wozniak2018}.
Making use of all supported mode orders plasmonic helices with one or more turns can be designed for resonating in a specific wavelength range and emitting with high directivity~\cite{Hoeflich2019,Kuen2024}. 
First attempts of co-integration have been successful, e.g.~single helices were coupled to a dipole source in the form of as a slit at the bottom~\cite{Wang2019subwavelength} and efficient coupling to surface-plasmon-polariton modes has been shown for the telecom wavelength regime~\cite{Wang2021}.
In both cases, also highly directional circularly polarized light emission was observed.

Here, we go a step further by designing and fabricating double helical antennas that work in the visible to telecom regime.
The two nano-wires couple, resulting in a hybridization of their fundamental plasmonic modes~\cite{Prodan2003a} to both symmetric (dipole-forbidden) and antisymmetric (dipole-allowed) modes~\cite{footnote1, Moradi2011}.
The antisymmetric mode concentrates the fields between the wires which allows both an enhanced coupling to dipolar quantum emitters placed in the gap and efficient interaction with circular polarized far-field radiation~\cite{Kuen2024}. 
This concept has the potential to realize solid-state ultrabright spin-selective single-photon sources.

Based on our established understanding of the excitation mechanisms in a single helix, we extended a semi-analytical design tool~\cite{Kuen2024} to describe double helices.
Full-field electromagnetic modeling refines the pre-selected design and provides insight to the antisymmetric mode patterns.
For our chosen design the wavelength range below 1000\,nm shows a chiroptical cross section almost twice as strong as for the single helical counterpart.
Then the helices are fabricated via direct electron beam writing using Au(acac)Me$_2$ as precursor and a subsequent oxygen plasma treatment to obtain a pure gold shell.
Finally, the fabricated antennas are optically characterized with respect to their circular dichroism.

\section{Results and Discussion}
\subsection{\label{1D}Theoretical Model}

\begin{figure}[ht]
\centering
\includegraphics[width=0.49\textwidth]{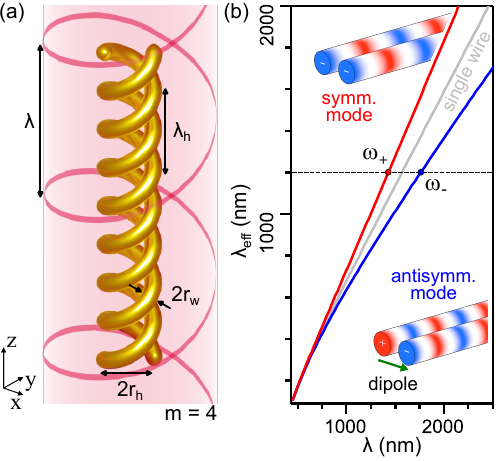}
\caption{\label{fig1}(a) Geometry of a helical plasmonic antenna. A double plasmonic right-handed helix of $m = 4$~turns, helix radius $r_h = 100\,$nm, helix pitch $\lambda_h = 430\,$nm and wire radius $r_w = 32\,$nm is illuminated by circularly polarized light propagating along the helix axis $z$. (b) Dispersion relations for both single and double straight wires showing the mode splitting due to the coupling of the plasmonic modes (using Equation~\eqref{eq:double_wire_shift}). Insets visualize the symmetric and antisymmetric modes on the thin straight wires.}
\end{figure}

\textbf{Figure~\ref{fig1}(a)} shows an artistic sketch of the studied system.
A double helix oriented along the $z$-axis is illuminated with a plane wave defined by a $k$-vector $k = 2\pi/\lambda$ pointing in the negative $z$ direction.
The right or left circular polarization (RCP and LCP) state is defined from the perspective of the receiver. For light propagating towards the observer, clockwise (counter-clockwise) rotation of the electric field vector is denoted as RCP (LCP)~\cite{Hecht2018}.
We quantify the power coupled from the wave to the helix modes based on the theoretical description for single helices as introduced earlier~\cite{Hoeflich2019}. 
The energy of the plane wave is given by $E = \hbar \omega = hc/\lambda = h k$ with the Planck constant $h$ and the vacuum speed of light $c$. 
The helix geometry is defined by the helix radius $r_h$, the pitch height $\lambda_h$ and the number of pitches $m$.
The two wires are coiled into a double helix, separated by $2\,r_h$ and have a circular cross section with radius $r_w$. 
On each of these helix wires Fabry-Perot modes of order $n$ can be excited~\cite{Novotny2007}.
The corresponding effective wavelength of the plasmonic Fabry-Perot modes $\lambda_{\text{eff}}$ can be calculated according to~\cite{Novotny2007} using the material parameters for a Drude model of gold.
Here, straight cylindrical wires of radius $r_w \ll \lambda$ are assumed, which has been proven to be a suitable approximation for wires curved in such a way that the helical turns are sufficiently separated to avoid near-field coupling~\cite{Hoeflich2019}.
Finally, the power transfer from plane wave to helix mode can be treated as a one-dimensional integral along the helix axis~\cite{Hoeflich2019,Kuen2024}.

In case of a double helix the modes will interact and hybridize (as depicted in Figure~\ref{fig1}(b)), depending on their distance $d=2r_h$, more exactly on their distance to wavelength ratio $d/\lambda_{\text{eff}} \propto k_{\text{eff}}\cdot d$ with $k_\text{eff} = 2\pi/\lambda_{\text{eff}}$ being the wire plasmon k-vector. 
Using the equations developed in~\cite{Moradi2011} which simplify for two cylindrical wires with identical radius, the two frequencies of the hybridized modes can be calculated as:

\begin{align}
    \omega_\pm^2 (k_\text{eff}) &= \omega^2 \pm \sqrt{\Delta} \label{eq:double_wire_shift} \\
    \text{with} \qquad \Delta &= \left [ \omega^2 K_0(k_\text{eff}\,d) \right ]^2 \frac{I_0(k_\text{eff}\,r_w)^2}{K_0(k_\text{eff}\,r_w)^2} \, , \label{eq:double_wire_shift_Delta}
\end{align}

where $\omega$ is the single wire plasmon frequency and $K_0$ and $I_0$ are the cylindrical Bessel functions used to describe the single wire mode field distribution. 
Keep in mind that both of these frequencies will lead to the same effective plasmon wavelength $\lambda_{\text{eff}}$, but for different hybridized modes (see Figure~\ref{fig1}(b)): 
(i)~The anti-symmetric mode with a positive charge opposed by a negative charge attracting each other over the gap leads to a decrease in the plasmon frequency and a red shift of the corresponding free space wavelength. 
(ii)~The symmetric mode where the coupling charge maxima are of equal sign and repel each other leads to an increase of the plasmon frequency.

Figure~\ref{fig1}(b) also explains why for a plane wave impinging along the $z$-axis only the antisymmetric mode is observed.  
The symmetric mode leads to no dipole moments perpendicular to the helix axis.
Therefore, from now on we only consider $\omega_-$, which leads to a red shift of the modes at the transition from a single helix to a double helix. 
The wire distance $d$ in \eqref{eq:double_wire_shift_Delta} is the distance between the charge density maxima of the standing wave. 
For the helix geometry this is in first order approximation identical to its diameter $d = 2r_h$.

With the double helix plasmon frequency $\omega_-$ calculated via \eqref{eq:double_wire_shift} both mode currents and overlap integral with the free space wavelength $\lambda$ can be calculated~\cite{Hoeflich2019,Kuen2024}.
An updated version of the one-dimensional modeling tool can be found online~\cite{1donline}.

\begin{figure*}[ht]
\centering
\includegraphics[width=0.99\textwidth]{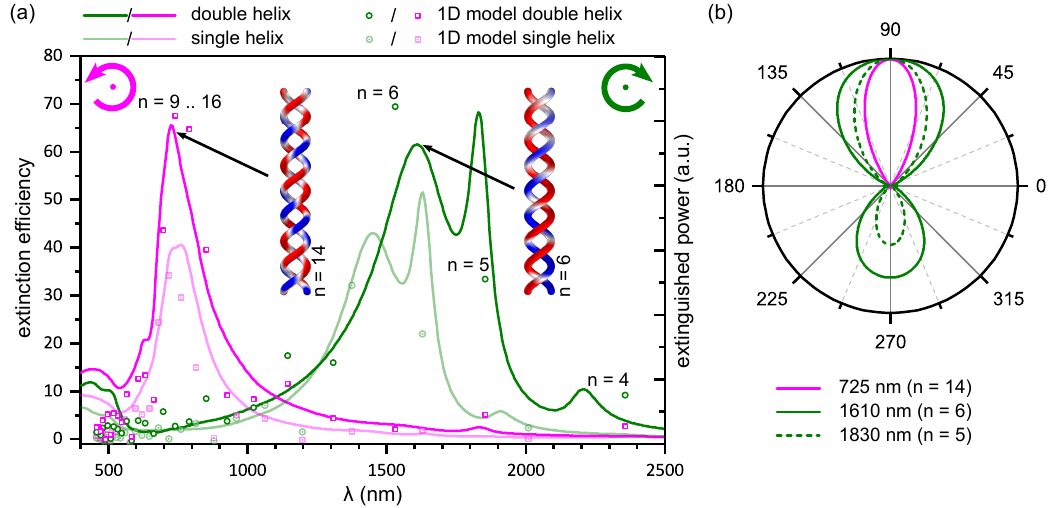}
\caption{\label{fig2}(a) Comparison of extinction efficiencies of free-standing single and double right-handed helices, when excited with RCP (green) and LCP (magenta) light. Subplots with normalized surface charge distribution of a high ($n = 14$) and a low order ($n = 6$) modes in the double helix show an antisymmetric pattern. Points represent position and power, coupled to the plasmonic modes according to the 1D theoretical model. (b) Numerically acquired normalized radiation pattern for scattering of the double helix at wavelengths of 725\,nm, 1610 and 1830\,nm for the respective handedness of strong chiroptical response.}
\end{figure*}

\subsection{\label{FFM}Full-Field Modeling}

Full-field electrodynamic modeling was used to study the position and width of the resonant modes of the double helix.
We characterize the spectral response by the extinction cross-section, with the extinguished light as the sum of scattering and absorption, normalized to the incident light intensity, and the geometric cross section of the helix projected on the plane of incidence~\cite{Bohren}.
\textbf{Figure~\ref{fig2}(a)} depicts the obtained spectra for a right-handed double helix with the material response of gold~\cite{Johnson1972} under LCP (magenta) and RCP (dark green) plane wave incidence.
The lighter colored graph is added for comparison with the corresponding single helix having identical geometry parameters and material response.
Additionally, the resonance positions and transferred powers obtained from the semi-analytical model are plotted for LCP (open circles) and RCP (open squares) in the same color scheme.
Surface charge distributions from the full-field simulations of selected modes are depicted in the insets of Figure~\ref{fig2}(b), with red/blue color indicating lack/excess of electrons.
The mode order $n$ is equal to the number of nodes of the fields inside the helix.

Overall, the observed spectral features of the double helix closely resemble those of the single helix, but red-shifted due to hybridization. The single helix resonance with $n = 5$ is therefore nearly exactly at the position of the double helix resonance with $n = 6$.
While in the low-energy region modes are efficiently excited for matching handedness of helix and incident light, in the high-energy region a strong response is obtained for opposite handedness~\cite{Hoeflich2019}. 
In the wavelength range above 1100\,nm the helix exhibits three distinct resonant modes for the excitation with light of matching handedness.
Compared to the single helix, these modes appear as red-shifted replica with increased extinction efficiency (see~Supporting Information (SI) for surface charge distributions of $n = 4 - 6$ modes).
This is well-described by the semi-analytical model, in which the resonance positions were calculated for the antisymmetric modes according to the dispersion relation in Figure~\ref{fig2}(a). According to Equation~\eqref{eq:double_wire_shift_Delta} the hybridization strength is proportional to $\lambda_\mathrm{eff}/d$, the ratio of single wire plasmon wavelength to wire distance. The decay length of the Mie mode's evanescent field outside the metallic cylinder depends on the wire plasmon wavelength.
Therefore, a smaller wavelength at a fixed wire distance leads to less interaction of the modes and, thus, less coupling and hybridization.

As another observation, the extinction efficiency is approximately doubled when transitioning from single to double helix. This can be directly understood from the semi-analytic model:
since the integration of the mode overlap must be carried out for two wires, the result will be doubled.
The only small change is the shift in the free-space wavelength to excite the anti-symmetric mode due to hybridization.
This shift is negligible at short wavelength, but even for the long wavelength resonances, both semi-analytical data points and full-field peak areas show a doubling of value and area, respectively.
This implies that the polarization selectivity of a double helix remains nearly the same as for the single helix.

The mode observed at the lowest energy is excited by RCP light at 2205\,nm. 
In this case, the number of nodes ($n = 4$) in the corresponding standing wave pattern equals the number of helix turns ($m = 4$).
While this mode is strongest for excitation with a localized dipole source in the near-field~\cite{Kuen2024}, the overlap condition with an external plane wave is non-optimal.
This is different for the dominating modes with matching handedness of the orders $n = 5$ and $n = 6$, as the mode overlap with the incident electric field is substantially enlarged~\cite{Hoeflich2019, Kuen2024, Novotny2007}.
The resonance width of the $n = 5$ mode at a wavelength of 1830\,nm is significantly reduced compared to the neighboring modes. 
This could potentially be interesting for an application as plasmonic cavity mode, as the reason for the reduced width is the decreased coupling to the far-field.
In plasmonic antennas this implies that all energy that is not scattered into the far-field is instead dissipated within the metal as heat.
Accordingly, the narrow $n = 5$ mode is absorption-dominated (see~SI for absorption and scattering cross-sections in Figure~S2) which limits realistic application scenarios. Still, due to its large slope such a mode can act as an extremely small and sensitive detector for chiral light.
In contrast, the $n = 6$ mode at 1610\,nm (in the telecom L-band) is strongly scattering and therefore well suited for antenna applications in the telecom range.
The inset depicts the corresponding standing wave pattern which is antisymmetric with respect to the opposing helix arms.
As expected, no symmetric surface charge distributions are excited.
However, a suitable beam of structured light can potentially excite these modes~\cite{Reich2020}.

For decreasing free space wavelengths from 1000 to 500\,nm the red-shift due to hybridization converges to zero (see Equation~\eqref{eq:double_wire_shift_Delta}), as the mode overlap between the wires scales with the effective plasmon wavelength. 
Between 600 and 900\,nm a multitude of closely spaced higher order modes is efficiently excited which we refer to as higher-order mode complex in the following.
The surface charge distribution for the spectral maximum at a wavelength of 725\,nm resembles an antisymmetric mode of order $n = 14$.
Furthermore, the mode complex is strongly scattering and correspondingly features low absorption (see SI for absorption and scattering cross-sections in Figure~S2).
Although the expected increase in absorption is obtained when approaching the visible range scattering is still strongly dominating. This is a consequence of the significant spectral distance of the resonance position to the interband transitions in gold.

In terms of possible applications, the far-field radiation properties of all excitable modes are important. 
Figure~\ref{fig2}(b) depicts far-field radiation patterns for the discussed modes of orders $n = 5,6$ as green dashed and solid lines and 14 as magenta line.
The $n = 5,6$ modes at 1830\,nm and 1610\,nm exhibit pronounced forward-scattering with moderate directivity. 
In contrast, the high-energy mode complex surrounding the $n = 14$ mode is exceptionally directional with a negligible portion of back-scattering.
The absolute value of the radiated intensity in forward direction is more than three times higher compared to the low-energy $n = 5,6$ modes (see SI: the non-normalized data in Figure~S2).
Together with the low absorption this offers potential applications for nonlinear light generation, where telecom light with one handedness is converted into visible light with the opposite handedness.
In case of second harmonic generation, for example, incoming light with a wavelength of 1400 - 1600\,nm is converted into light with a wavelength of 700 - 800\,nm and efficiently re-radiated into the far-field, as both bands feature highly efficient modes.

\subsection{\label{sec:exp}Experimental Realization and Characterization}

\begin{figure*}[htb!]
\centering
\includegraphics[width=0.49\textwidth]{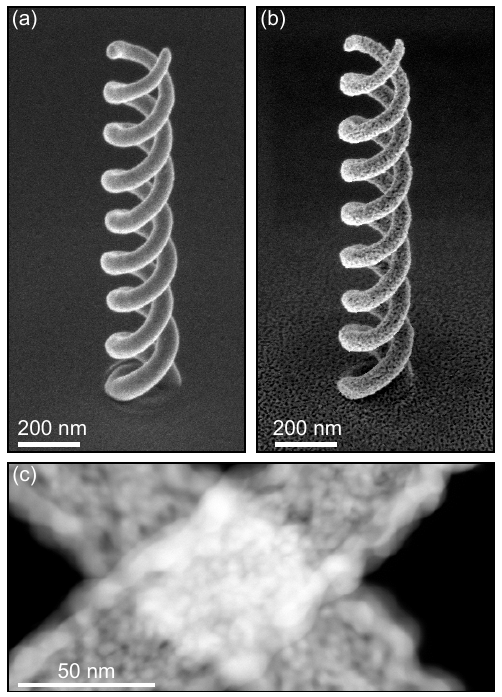}
\caption{\label{fig3}(a) SE micrograph of the as-deposited double helix made from Au, 45\textdegree~side view. (b) SE micrograph of the double helix after oxygen plasma cleaning, 45\textdegree~side view. (c) STEM HAADF image of double helix after purification depicting the presence of gold shell.}
\end{figure*}

Double helices were fabricated by direct electron beam writing~\cite{Utke2008, Hoeflich2011} using Au(acac)Me$_2$ as precursor compound.
The gaseous precursor molecules adsorb on the surface and are locally dissociated by focused electron beam impact~\cite{Utke2022}. Due to its direct nature, the technique is ideal for printing complex nanoscale geometries~\cite{Winkler2019} and chiral structures in particular~\cite{Sanz-Hernandez2020, Esposito2015}.

\textbf{Figure~\ref{fig3}}(a) depicts the SE micrograph of the resulting double helical geometry.
During deposition, the electron beam jumps back and forth on a circular path, alternating between the two helix arms with a slight counterclockwise (clockwise) shift.
Double helices of either handedness have been fabricated to cross-check the geometry-dependent sign-change of the chiroptical response.

The beam path parameters to be optimized are the distance between two neighboring points on the same helix arm and dwell times for each point, as these determine the tangential beam velocity and thus the vertical growth rate.
%Here, the beam path combines two timescales.
%The dwell times and pixel distance alongy the circular path define the beam velocity for the vertical structures growth,
On a much shorter time scale the radial movement provides for quasi-parallel printing and assists in precursor refreshment~\cite{Keller2018}.  
%With increasing height the wire-like geometries are less effective in dissipating the input energy as heat into the substrate~\cite{Utke2002}. 
With increasing structure height, the quasi-continuous energy input by the electron beam at the structure apex increases the local temperature on the deposit surface~\cite{Utke2002, Tsarapkin2024}.
The altered diffusion kinetics and the thermally enhanced molecule desorption result in decreasing adsorbate density  with increasing structure height such that the deposition takes place in the adsorbate-limited (mass transport-limited) regime~\cite{Sanz-Hernandez2020, Mutunga2019}. 
The resulting nonlinear decrease in the vertical growth rate was compensated by a dynamic increase of the dwell times.

In addition to the beam path parameters, the electron beam current determines the number of electrons available in the focal spot. In the mass-transport-limited regime with continuously decreasing precursor availability small beam currents are preferred~\cite{Keller2018, Fowlkes2018}. We therefore used an electron beam current of 100\,pA.

Finally, the primary energy of the electrons determines the size of the collision cascade, which is related to the yield and spatial distribution of secondary electrons that drive the process~\cite{Silvis2002}. 
High beam energies above 15\,keV, while beneficial in allowing the smallest electron beam spots and larger depth of field, result in elongation of the wire cross-section that is more pronounced the more horizontal they are~\cite{Fowlkes2018}. 
In this work, a beam energy of 5\,kV was used, resulting in a relatively round wire shape.
The optimum position of focal plane during deposition was fixed above the sample surface by an auxiliary deposit for focusing.
The reduced size of the collision cascade also reduces the so-called halo, a faint, unwanted deposit that covers an area up to a few micrometers away from the actual deposit and is caused by backward-scattered electrons from the substrate~\cite{Hoeflich2017} as well as by forward-scattered electrons from larger structure heights~\cite{Tsarapkin2024}.
These parasitic electrons pin-down precursor molecules around the helical structure what efficiently prevents spurious deposition inside the helical structure base due to electron beam toggling between the two helix arms~\cite{Utke2008, Fowlkes2018, Szkudlarek2014}. 
It is important to note that the minimum achievable dimensions of the double helix are not given by the wire thickness, but by attractive Coulomb forces.
As soon as the helix arms come too close to each other, they attract each other due to the constant supply of electrons during writing and eventually touch.

The deposited material that forms the helical structure consists of single-crystalline gold particles embedded in a carbonaceous matrix~\cite{Hoeflich2011}.
The nano-granular structure of the deposit is still frequent in direct electron beam writing~\cite{Huth2018}, although strategies for higher metal content FEBID material nanoprinting are actively researched~\cite{Barth2020, Reisecker2024}.
As the cleavage of the mostly organometallic precursor molecules is not bond-selective, a significant amount of the ligand material and residual hydrocarbons are often incorporated~\cite{Utke2022}.
The desired plasmonic material response, however, requires a pure metal surface with a thickness exceeding the penetration depth of light in the relevant spectral range. 
This is achieved by purification using an ozone plasma treatment at room temperature~\cite{Haverkamp2016}.
The result of which can be seen in the scanning electron micrograph in Figure~\ref{fig3}(b) as a roughened surface due to the remaining gold particles.
If the wire diameter is too thin, plasma purification leads to bending of the helix arms, which eventually destroys the helix. 
This sets a lower limit for the wire diameter of 60\,nm.
After purification the helices were characterized using scanning transmission electron microscopy (STEM) for their microstructure and composition.
Figure~\ref{fig3}(c) shows the corresponding high-angle annular dark-field (STEM-HAADF) image of the two helix arms after purification in which the agglomerated gold particles at the surface form a thin layer of approximately 15\,nm.
The core of the helix resembles the typical granular structure of gold nanocrystals surrounded by carbon.

\begin{figure}[htb!]
\centering
\includegraphics[width = 0.49\linewidth]{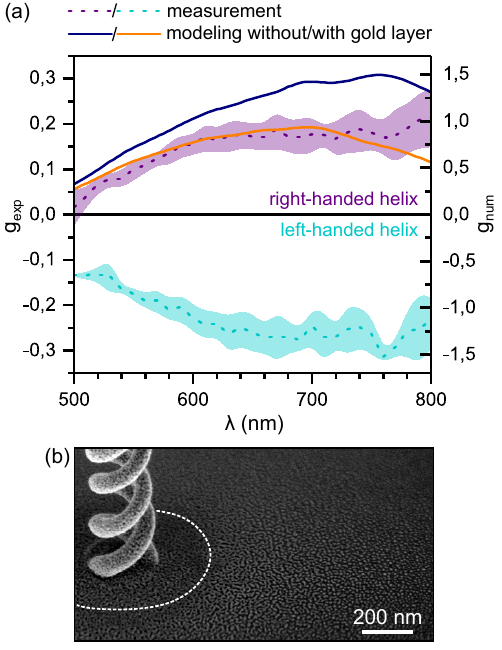}
\caption{\label{fig4} (a) Comparison of the experimental dissymmetry factor $g_{\text{exp}}$ of right-handed (navy) and left-handed (cyan) helices with theoretical $g_{\text{num}}$ obtained by full-field modeling of the transmission setup with the right-handed double helix on the substrate. (b) SE micrograph of the double helix after oxygen plasma purification 45\textdegree~side view. Spurious gold arising from halo deposition is visible on the surface and forms a quasi-continuous layer in the vicinity of the helix base, highlighted by a dashed line.}
\end{figure}

Optical transmission spectroscopy was employed for the characterization of the chiroptical response of the double helical antennas.
The measurements were carried out at the level of the individual nanoantennas using a high numerical aperture (NA) refractive objective for focusing the light onto the sample. 
The size of the focus was chosen to guarantee homogeneous illumination while maximizing the signal-to-noise-ratio. 
The scattered light, along with the primary excitation beam transmitted through the double helix, is collected using a long working distance objective and directed to a photodiode (see Methods section for more details). 
The dissymmetry factor g~\cite{Hoeflich2019, Kuhn1930} can be reformulated from the numerically accessible extinction cross-section to transmittance measured in the experiment:
\begin{equation}
g_{\text{num}} = \frac{2(\sigma_{ext}^{\text{LCP}} - \sigma_{ext}^{\text{RCP}})}{\sigma_{ext}^{\text{LCP}} + \sigma_{ext}^{\text{RCP}}} = \frac{2(T_{\text{RCP}} - T_{\text{LCP}})}{2- T_{\text{RCP}} - T_{\text{LCP}}} = g_{\text{exp}}\, .
\end{equation}
This provides a direct comparison of the experiment to numerical modeling (see Methods Section for more details).
\textbf{Figure~\ref{fig4}(a)} displays the experimentally obtained dissymmetry spectra in the range of 500 to 800\,nm for left-handed (cyan) and right-handed (navy) double helices compared to the full-field modeling.
When taking into account the residual gold on the surface around each helix caused by halo deposition, experiment and modeling show good agreement. Figure~\ref{fig4}(b) depicts a close up of the surface region around the helical antenna covered by gold nanoparticles that coalesce close to the antenna base. Here, it has to be pointed out that this residual gold does not change the functional behavior or the overall spectral response of the double helix, but only slightly attenuates the chiral response due to the presence of achiral scatterers/absorbers on the surface (see SI for more details on the modeling setup).
As expected from modeling, in this wavelength range the individual modes of the high-energy mode complex cannot be spectrally resolved.
Such a broadband operation with dissymmetry factors between 0.2 and 0.3 is interesting for the coupling to multicolor quantum emitters in the visible range~\cite{Kuen2024, Esmann2024}.

\newpage
\section{Conclusions}

Here, plasmonic double helices acting as sensitive antennas for circularly polarized light were designed and experimentally demonstrated.
Upon interaction with circularly left and right polarized light, modes with antisymmetric surface charge distribution along the coupled wires are efficiently excited.
The previously developed analytical tool for a single helix~\cite{Hoeflich2019} was extended to account for the second helical arm by implementing mode coupling~\cite{Moradi2011}. 
Compared to single helices the optical cross sections are doubled.
The modes can efficiently be tuned by adjusting the antenna geometry.
The wire distance represents an additional design parameter, which can be used to adjust resonances by the degree of hybridization between the two wire modes.

The designed helices were modeled using full-field finite element and finite-difference time-domain simulations to retrieve the width and surface charge distributions of their modes.
The calculated far-field radiation especially for the high-energy mode complex is highly directional.
The experimental realization of the double helical antennas was carried out using direct electron beam writing followed by ozone plasma purification~\cite{Haverkamp2016}.
The fabricated double helices were characterized with respect to their transmission, revealing large circular dissymmetry factors in a broad wavelength range between 500 and 800\,nm.

The here introduced double helical antennas have several promising applications. 
First, they are a possible building block for chiral metamaterials or phase arrays~\cite{Esposito2015, Kaschke2015, Kravets2018, Khaliq2023} with the prospect of two-dimensional chiroptical components for (color-sensitive) beam steering and focusing.
In this regard, the increased optical cross section allows for smaller antenna heights and therefore thinner and more stable devices while allowing for better integration with far-field optics.

Second, individual double helical antennas can serve as highly directional far-field receivers concentrating chiral light below the diffraction limit.
The spectral properties of double helices could be enhanced further by embedding them into complex dielectric environments or coupling their emission to e.g.~photonic crystal waveguides~\cite{DeAngelis2010,Zhang2014}.
The special design would even enable a chip-based non-linear helicity switch in which the circularly polarized input light generates second harmonic radiation with twice the energy and opposite circular polarization.

Third, adding the second wire to a helical design allows to confine the evanescent plasmonic fields to the space between the wires, enabling strong light-matter interaction at the nanoscale.
This is particularly favorable for coupling emitters with in-plane dipolar moments, such as e.g.~quantum emitters in 2D materials, directly at the helix base to enhance and steer their emission while imparting a well-defined spin to the emitted photons~\cite{Kuen2024}.
When used as a scanning probe in a cantilever-based sensing technique such as scattering near-field optical microscopy (SNOM), the helical antennas can be coupled to more than one emitter subsequently~\cite{Taminiau2008, Farahani2005} or map out the circularly polarized photoluminescence of near-IR emitters such as carbon nanotubes~\cite{Reich2004} with sub-wavelength resolution.
The fabrication of such sophisticated probes is only possible by direct electron beam writing, which allows the antennas to be fabricated directly on a non-planar substrate such as a cantilever~\cite{Kusch2024}.
Finally, double helical antennas may also open new possibilities of active plasmonics by harnessing optomechanical effects, as charging will lead to a height dependent change in wire distance\cite{chen_electromechanically_2016}.
Hence, in a broader sense, these results can form the basis for future developments in the fields of plasmonic circuitry and photonic quantum technology.

\section{Experimental Section}
\threesubsection{Numerical Modeling}\\

Full-field modeling was carried out using commercial Maxwell solvers.
Two different modeling techniques were employed.
The scenario without substrate corresponding to the described case in the semi-analytical model was modeled using the finite element method (FEM). 
The scenario with substrate was modeled using a finite-difference time-domain (FDTD) technique.
While in FEM the triangular meshing avoids artifacts in the local field distributions at curved surfaces and thus provides a good representation of surface charge distributions, the FDTD with the light propagating in discrete time steps and measurement monitors provides a direct representation of the experiment.

\threesubsection{Sample Fabrication}\\

High precision glass cover slides (Roth) with a thickness of $170\pm5$\,\textmu m were used as transparent substrates. 
To obtain a conductive sample surface, a thin layer of ITO was deposited by sputter coating (AJA International Inc.).
Thickness and complex refractive wavelength of the ITO layer were determined by ellipsometry measurements of a silicon reference chip covered in the same sputter run.
The obtained ITO layer thickness was 30\,nm with a weakly dispersive refractive index around 2.0 and negligible losses in the investigated wavelength range. 
The helices were fabricated by direct electron beam writing in a Thermo Fisher Helios 5 UX dual beam microscope using the metal-organic precursor Au(acac)Me$_2$ \\(dimethyl\-(acetyl\-acetonate)\-gold(III)).
The precursor reservoir is heated to 30~\textdegree C to mobilize and locally deliver the gas molecules by an injection nozzle.
The electron beam locally dissociates the physisorbed molecules, forming a deposit from non-volatile parts after decomposition. 
The base background pressure before deposition was 5.3$\cdot$10$^{-7}$~mbar.
Upon opening the nozzle valve, the background pressure stabilized \\ at 5.8$\cdot$10$^{-7}$~mbar.
Deposition was started few minutes later after recovering the pressure close to the initial value to ensure for uniform precursor delivery over time.
The tangential pitch between two points in the same helical arm was chosen to be 0.04\,nm, with dwell times varying from 200 to 700\,\textmu s according to the  nonlinear calibration function.
The resulting helices were then purified in a Zepto Diener oxygen plasma cleaner for 60 sec with 0.2\,sccm oxygen flux which roughly corresponds to less then 0.3\,mbar background pressure to obtain closed pure gold shell.

\newpage
\threesubsection{Characterization}\\
For the characterization of the microstructure and composition of the helices a probe-corrected ThermoFisher Scientific Titan Themis 200 G3 transmission electron microscope at an acceleration voltage of 200\,kV was used.
The optical characterization was carried out using a optical transmission microscopy setup in the wavelength range between 500--800\,nm.
Circularly polarized light from a super-continuum laser (FIR-20, NKT Photonics) in combination with spectral filter (VIS HP8, NKT Photonics) was focused onto the backside of the sample to a waist diameter of approximately 1~\textmu m for all wavelengths using a high NA refractive objective (0.95 NA, CFI Plan Apo Lambda 60XC, Nikon).
RCP and LCP light are generated by combining a linear polarizer with a broadband achromatic quarter-wave plate (AQWP10M-980, Thorlabs).
The transmitted signal of one particular helix was collected using a second refractive objective with 100x magnification and NA = 0.7 (MY100X-806, Mitutoyo) and measured using a photodiode (S120C, Thorlabs) connected to an optical power meter (PM5020, Thorlabs).
The same measurement is performed on a substrate position without any structure, allowing for the normalization of the helix transmission.

%\medskip

% \textbf{Supporting Information} \par 
% More details on the semi-analytical model, full-field modeling, and characterization of the double helical antennas can be found in the Supporting Information.
% Supporting Information is available from the Wiley Online Library or from the author.

\medskip
\textbf{Acknowledgements} \par 
The authors would like to thank Sabrina Jürgensen and Stephanie Reich for fruitful discussions about optical characterization.
This work was funded by the Deutsche Forschungsgemeinschaft (DFG, German Research Foundation) under Project ID HO5461/3-1 “chiralFEBID” and and by the Swiss National Fund by the COST -- SNF project IZCOZ0\_205450.
The authors want to furthermore acknowledge financial support by the EU COST action CA 19140 ‘FIT4NANO’ (www.fit4nano.eu).

\medskip
\textbf{Conflict of Interest} \par 
The authors declare no conflict of interest.

\medskip

\newpage
\section*{Supporting Information}

\setcounter{section}{0}
\setcounter{figure}{0}
\renewcommand{\figurename}{Figure}
\renewcommand{\thefigure}{S\arabic{figure}}

\section{Extension of the Semi-Analytical Model}

\begin{figure}[htb]
     \centering
     \includegraphics[width=0.95\linewidth]{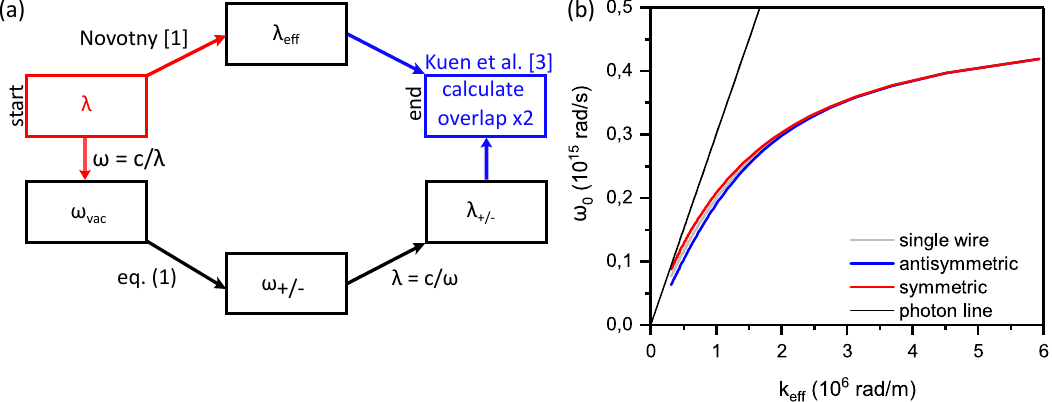}
     \caption{(a) The calculations needed to retrieve the 1D-overlap integration. (b) Dispersion relations for both single and double straight wires showing the mode splitting due to the coupling of the plasmonic modes (using Equation 1 in the main manuscript).}
     \label{fig:SI1}
\end{figure}

To perform a one-dimensional overlap integral between free space excitation circularly polarized light and a plasmonic wire Fabry-Perot mode, the involved wavelengths have to be calculated. 
This is done as sketched in the flowchart depicted in Figure~\ref{fig:SI1}(a), starting from the free space wavelength $\lambda$. 
According to~\cite{Novotny2007}, the effective wavelength of the plasmons $\lambda_\text{eff}$ is calculated for a straight wire with the given helix wire radius $r$. 
The metal parameters for gold in a simple Drude-Sommerfeld model with $\varepsilon_\infty = 1.54$ and $\lambda_\text{plasma} = 143\cdot 10^{-9}$ were taken from~\cite{Etchegoin2006}.
However, since we are working with a double helix, the coupling between the two wires comprising the helix has to taken into account. 
This is done via Equation~(1) in the main manuscript, using the free space frequency $\omega=c/\lambda$ with $c$ being the vacuum speed of light.
The distance between the wires was chosen to be twice the radius of the helix, since the coupling occurs between the charge density extrema, which are always at the same $z$-position along the helix axis.
The resulting new frequencies $\omega_\pm$ are then converted back to corresponding excitation wavelengths $\lambda_\pm$, used identically as in Kuen \textit{et al.}~\cite{Kuen2024}.
As the double helix consists of two identical wires the resulting extinction has to be multiplied by a factor of two.
Figure~\ref{fig:SI1}(b) shows the dispersion relations comparison of the single and double wires (cf. Figure~1(b) in the main manuscript) in the recalculated units $\omega$ vs. $k_{eff}$.
The version of the code used to calculate the data within this paper can be found here: \\
\url{https://sourceforge.net/projects/plasmonic-helix-1dmodel/files/plasmonic_helix_v1_3_etchegoin.ipynb/download}

\section{Numerical Modeling}

As described in the main manuscript, full-field modeling was carried out using commercial Maxwell solvers.
Two different modeling techniques were employed.
The scenario without substrate corresponding to the described case in the semi-analytical model was modeled using the finite element method (FEM). 
The scenario with substrate was modeled using a finite-difference time-domain (FDTD) technique.
In FEM (COMSOL Multiphysics) the helix geometry and plane wave illumination was implemented into a full-field scattered-field environment terminated by perfectly matched layers (PML).
The normal component of the dielectric displacement with respect to the helix surfaces was used to visualize the surface charge distribution.
FDTD modeling (Ansys Lumerical FDTD) of the same geometry and using the same material parameters proved that the substrate influence is negligible for the aspect ratio of this geometry.
The intensities of total and scattered field were obtained by projecting the corresponding Poynting vectors onto a surrounding integration surface and used to obtain the scattered and absorbed (total - scattered) intensities.
The antenna geometry consist of two helices with a relative rotation of 180\textdegree~with respect to each other. 
The helices have $m = 4$ turns with a helix pitch $\lambda_h$ of 430\,nm, a helix radius $r_h$ of 100\,nm and a wire radius $r_w$ of 32\,nm with the wires being terminated with round caps of the same radius.
The permittivity of gold was defined using an analytical fit function to the data of Johnson and Christie~\cite{Johnson1972} in both cases.
It should be noted that the models do not take into account the thickness of the gold shell and random gold particles that are scattered at the base of the helix.
These particles are randomly deposited with secondary and scattered electrons and become visible after cleaning with oxygen plasma.

The numerical dissymmetry factor $g_{\text{num}}$ was obtained by modeling the transmission of LCP and RCP Gaussian beams through the right-handed double helix using the FDTD method and calculated by using the same formula as for $g_{\text{exp}}$ in Equation~3 of the main manuscript.
The transmissions of RCP and LCP light were normalized to the transmission of an identical system without the helix.
The transmission values were extracted using the "farfield3d" script command, which excludes near-field effects and allows to put the transmission monitor closer to the helix.
Only the field vectors within a NA = 0.7 were used to calculate the transmission, reflecting the parameters of the collection objective used in the actual experiment.

During helix fabrication, a thin deposit layer is usually co-deposited at the helix base due to secondary and backscattered electrons.
After oxygen plasma cleaning, it forms an almost solid gold disk that electrically connects the two helical arms.
Accordingly, we conducted additional full-field modeling of light transmission through the system with a thin disk of 8~nm thickness and 700~nm radius.
The results are shown as a solid orange line in Figure~3(b) of the main manuscript.
Effectively, the presence of the disk reduces the dissymmetry factor and blue-shifts its maxima, resulting in a better fit to the experimental results.

In the following we present additional plots with supplementary numerical investigation details.

\begin{figure}[htb!]
\centering
\includegraphics[width=0.95\linewidth]{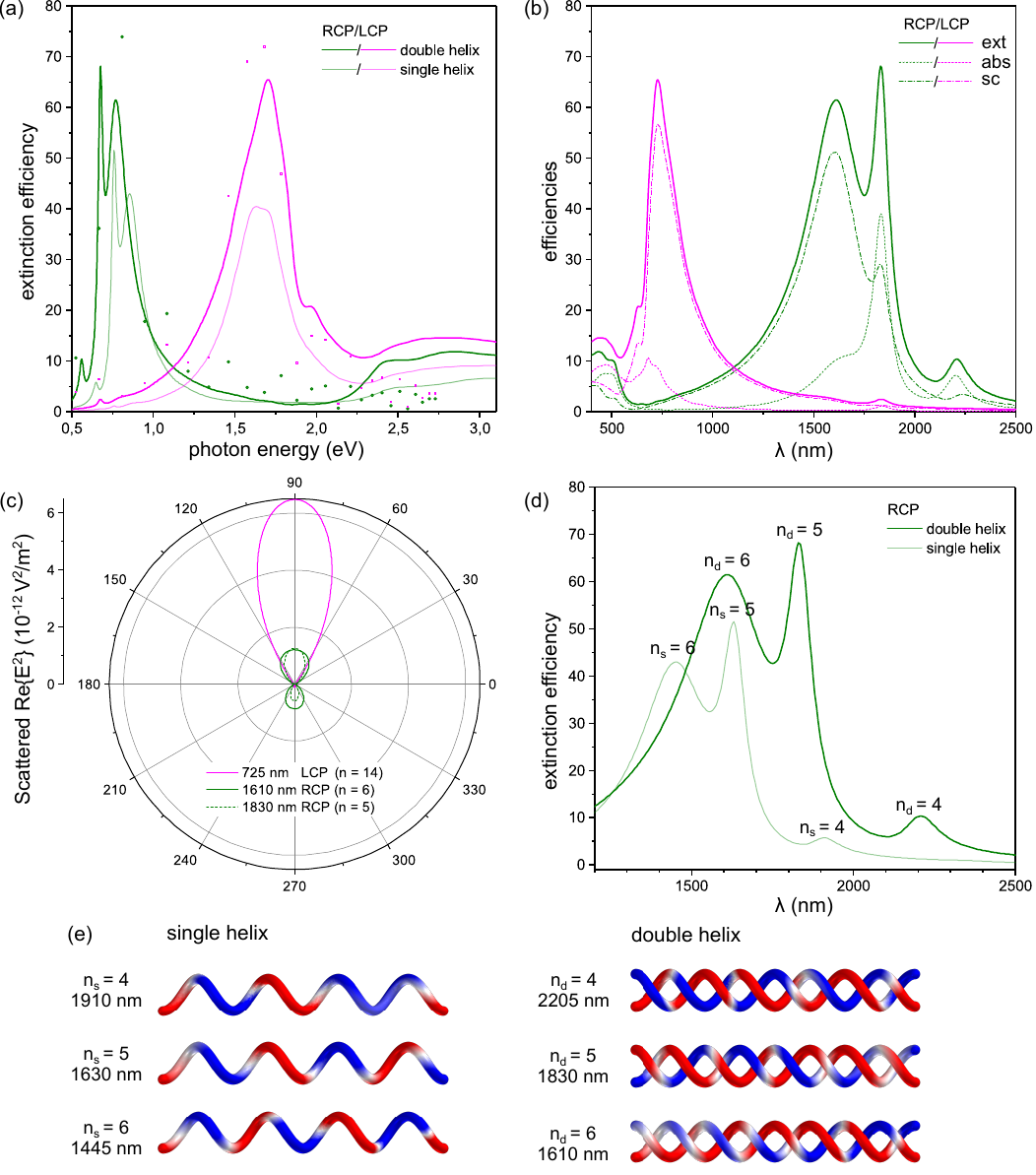}
\caption{\label{fig:SI2} (a) FEM simulation results on total extinction efficiency for the double helix plotted vs. photon energy, investigated in the main manuscript. (b) FEM simulation results on total scattering and absorption for the double helix, investigated in the main manuscript. (c) Radiation pattern based on FDTD simulation results on scattering of the double helix at 725\,nm LCP and 1610 and 1830\,nm RCP in the far-field. (d) FEM simulation results revealing the modes $n = 4-6$  positions and the respective (e) surface charge distribution in single and double helices upon excitation with RCP light.}
\end{figure} 

Figure~\ref{fig:SI2}(a) displays the same data as Figure~2(a) in the main manuscript but plotted over the energy.
Here, the Fabry-Perot modes from the semi-analytical model appear spectrally well-separated in the range of 0.5 - 2.5\,eV.
Again, a pronounced shift due to plasmon hybridization at the double wires is only visible in the low-energy region. 
The resonance broadening obtained from the full-field modeling without substrate  is generally larger for the higher order (higher energy) modes than that for the lower order modes.
Therefore, the high-order modes in the modeled spectra are not spectrally resolvable in the energy scale either.

Figure~\ref{fig:SI2}(b) disentangles the contributions of scattering and absorption to the total extinction.
In addition to the extinction cross-section, the absorption cross-sections $\sigma_\text{abs}$ are shown as dashed lines and the scattering cross-sections $\sigma_\text{sca}$ as dashed-dotted lines for the respective handedness of the incident light.
While the two modes discussed in the main manuscript ($n=6$ and $n=14$) are mainly scattering, the sharp $n=5$ mode is mostly absorbing. 

Figure~\ref{fig:SI2}(c) displays the radiation patterns of the free standing double helix as a scatterer as in Figure~3(a) of the main manuscript. 
Here, the data is not normalized to visualize the absolute values of radiated power for the discussed modes.
The considered modes are the $n = 5, 6$ modes at 1830\,nm 1610\,nm respectively for right circularly polarized light and the mode $n = 14$ at 725\,nm for left circularly polarized light.
In all cases, the double helix exhibits directional scattering characteristics, with the maximum power scattered within approximately 60\textdegree.
Interestingly, the radiation pattern at 725\,nm is exceptionally asymmetric, with the scattered light propagating along the helical axis in the same direction as the incident light.
This can be explained by the fact that the wavelength is almost twice as short as the height of the helix, which results in retardation effects and excitation of several neighboring modes.
This is diminished at 1610 and 1830\,nm, at which the helix scatters some radiation backwards.
In consequence the absolute intensity that is radiated in forward-direction is more than three times larger for the high-energy mode complex at 725\,nm compared to its lower energy counterparts.

Finally, Figure~\ref{fig:SI2}(d) presents a close-up comparison of the modes $n = 4, 5, 6$ position for the single and double helix of the same geometry, as shown in Figure 2(a) of the main manuscript.
As predicted by the semi-analytical 1D model, the modes of the same order appear red-shifted for the double helix.
In addition, Figure~\ref{fig:SI2}(e) provides a detailed insight into the surface charge distribution of the selected modes for both the single and double helix.

\section{Optical Measurements}

\begin{figure}[htb!]
\centering
\includegraphics[width=0.5\linewidth]{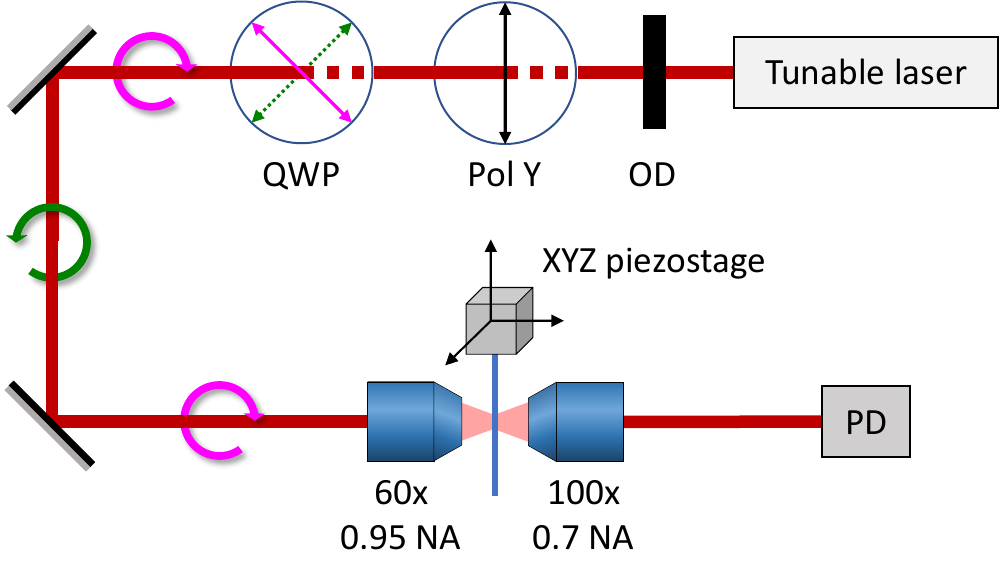}
\caption{\label{SI:char} Sketch of the transmission optical microscopy setup. OD: optical density filter; Pol Y: polarizer in $y$-direction; QWP: quarter-wave plate; PD: photodiode.}
\end{figure}

The fabricated helices were optically characterized in a transmission setup sketched in Figure~\ref{SI:char}.
As the excitation source a super-continuum laser (FIR-20, NKT Photonics) was used in combination with a spectral filter (VIS HP8, NKT Photonics) to prepare narrow spectral lines of width less then 2.5\,nm, in the spectral range of 500 to 800\,nm. 
This laser light was focused onto the substrate to a spot with a diameter of approximately 1~\textmu m using a 60x magnification objective with a large numerical aperture (0.95~NA, CFI Plan Apo Lambda 60XC, Nikon). 
To achieve this the back aperture of the objective was only partially illuminated.
RCP and LCP light were generated by combining a linear polarizer and an achromatic quarter-wave plate (AQWP10M-980, Thorlabs). 
Both the scattered light together with the undisturbed primary excitation beam passing a double helix were then collected using a long working-distance objective, with 100x magnification and NA = 0.7 (MY100X-806, Mitutoyo), and subsequently directed to a photodiode (S120C, Thorlabs) connected to an optical power meter (PM5020, Thorlabs).
The same measurement was performed for each wavelength at a spot without any structure to allow normalization of the helix transmission which ultimately gives a measure for the extinction.

\bibliographystyle{MSP}
\bibliography{MSP-template}

\end{document}